\documentclass[%
reprint,
superscriptaddress,
showpacs,
nofootinbib,
nobibnotes,
 amsmath,amssymb,
 aps,
prc,
floatfix,
]{revtex4-1}

\usepackage{graphicx}
\usepackage{dcolumn}
\newcolumntype{.}{D{.}{.}{-1}}
\usepackage{bm}
\usepackage{float,array,booktabs,amsmath,amssymb}

\begin{document}

\preprint{APS/123-QED}

\title{\boldmath 
Investigation of pair-correlated $0^+$ states in $^{134}$Ba via the $^{136}$Ba($p,t$) reaction}
\author{J.\,C.~Nzobadila~Ondze}
\affiliation{Department of Physics and Astronomy, University of the Western Cape, P/B X17, Bellville 7535, South Africa}%
\author{B.\,M.~Rebeiro}
\altaffiliation[Present address: ]{Univ Lyon, Univ Claude Bernard Lyon 1, CNRS/IN2P3, IP2I Lyon, UMR 5822, F-69622, Villeurbanne, France}
\affiliation{Department of Physics and Astronomy, University of the Western Cape, P/B X17, Bellville 7535, South Africa}%
\author{S.~Triambak}
\email{striambak@uwc.ac.za}
\affiliation{Department of Physics and Astronomy, University of the Western Cape, P/B X17, Bellville 7535, South Africa}%
\author{L.~Atar}
\affiliation{Department of Physics, University of Guelph, Guelph, Ontario N1G 2W1, Canada}%
\author{G.\,C.~Ball}
\affiliation{TRIUMF, 4004 Wesbrook Mall, Vancouver, British Columbia V6T 2A3, Canada.}
\author{V.~Bildstein}
\affiliation{Department of Physics, University of Guelph, Guelph, Ontario N1G 2W1, Canada}%
\author{C.~Burbadge}
\affiliation{Department of Physics, University of Guelph, Guelph, Ontario N1G 2W1, Canada}%
\author{A.~Diaz Varela}
\affiliation{Department of Physics, University of Guelph, Guelph, Ontario N1G 2W1, Canada}%
\author{T.~Faestermann }
\affiliation{Physik Department, Technische Universit\"{a}t M\"{u}nchen, D-85748 Garching, Germany}%
\author{P.\,E.~Garrett}
\affiliation{Department of Physics, University of Guelph, Guelph, Ontario N1G 2W1, Canada}%
\affiliation{Department of Physics and Astronomy, University of the Western Cape, P/B X17, Bellville 7535, South Africa}%
\author{R.~Hertenberger}
\affiliation{Fakult\"{a}t f\"{u}r Physik, Ludwig-Maximilians-Universit\"{a}t M\"{u}nchen, D-85748 Garching, Germany}%
\author{M.~Kamil}
\affiliation{Department of Physics and Astronomy, University of the Western Cape, P/B X17, Bellville 7535, South Africa}%
\author{R.~Lindsay}
\affiliation{Department of Physics and Astronomy, University of the Western Cape, P/B X17, Bellville 7535, South Africa}%
\author{J.\,N.~Orce}
\affiliation{Department of Physics and Astronomy, University of the Western Cape, P/B X17, Bellville 7535, South Africa}%
\author{A.~Radich}
\affiliation{Department of Physics, University of Guelph, Guelph, Ontario N1G 2W1, Canada}%
\author{H.\,-F.~Wirth}
\affiliation{Fakult\"{a}t f\"{u}r Physik, Ludwig-Maximilians-Universit\"{a}t M\"{u}nchen, D-85748 Garching, Germany}%
%
%
%
 
\date{\today}
%
 \begin{abstract}
 We performed a high resolution study of $0^{+}$ states in $^{134}$Ba using the $^{136}$Ba($p,t$) two-neutron transfer reaction. Our experiment shows a significant portion of the $L = 0$ pair-transfer strength concentrated at excited $0^+$ levels in $^{134}$Ba.
Potential implications in the context of $^{136}$Xe $\to$ $^{136}$Ba neutrinoless double beta decay matrix element calculations are briefly discussed.   
\end{abstract}

\maketitle
\noindent
The even-mass barium isotopic chain ($Z = 56$) is a fertile testing ground for nuclear structure models and also important for nuclear astrophysics studies and tests of fundamental symmetries. 
For example, theory calculations predict an enhanced octupole collectivity around $^{112}$Ba, which is located on the $N = Z$ line~\cite{Skalski,Heenen}. Possible octupole correlations have been observed in the neutron-deficient $^{118}$Ba~\cite{Smith} and $^{124}$Ba~\cite{Mason} isotopes, while there is clear experimental evidence of octupole deformation in the ground states of neutron-rich Ba nuclei around $N = 88$~\cite{bucher1,bucher2}. Furthermore, for $N \le 82$, in the $A \sim 130$ region, the even Ba isotopes are expected to be shape transitional~\cite{marshalek}. Their shape evolution ranges from nearly spherical (semi-magic at $N = 82$) to $\gamma$-soft~\cite{casten1}, where the shape changes are characterized by quantum phase transitions (QPTs)~\cite{QPT,nomura}. Within the interacting boson model (IBM), the $^{134}$Ba isotope was identified as a potential $E(5)$ symmetry critical point~\cite{casten2} for a second-order QPT. 
Independently, from a nuclear astrophysics persective, fractional abundance ratios of odd-to-even barium isotopes as well as relative elemental ratios such as [Ba/Fe] and [Ba/Eu] etc, offer insight into the $r$ and $s$-process neutron capture contributions~\cite{star1,Takuji} to heavy element nucleosynthesis in stellar environments. This is particularly relevant in metal-poor stars, where the dominant contribution to elemental abundances is expected to be from the $r$-process~\cite{Truran}. One interesting example is the subgiant HD140283, in which case the fractional barium abundances obtained from independent spectral analyses have yielded inconsistent results. These discrepancies have stirred a debate on the assumed $r$-process origin of the odd Ba isotopes during the early stages of galactic evolution~\cite{star2,Magain:95,francois,collet,Lambert}.

In the context of fundamental symmetries, $^{136}$Ba is the daughter nucleus in $^{136}$Xe double beta decay, an attractive candidate to search for the lepton-number-violating neutrinoless double beta ($0 \nu 2\beta$) transformation. Observation of such decays would unequivocally show that the neutrinos are their own antiparticles (i.e. they are Majorana fermions). In such a scenario, other useful information on lepton-number-violating new physics or the absolute neutrino mass scale, etc. can only be obtained if the nuclear matrix element (NME) for the decay is accurately known~\cite{Engel:2017,ejiri}.  
To this end, a
variety of many-body techniques are used to evaluate $0\nu2\beta$ decay NMEs in several candidate nuclei. These calculations have yielded significantly discrepant results for individual cases~\cite{Engel:2017}. Addressing this apparent model dependence in all double beta decaying nuclei remains an important issue. We focus on this aspect here. 

The dominant contribution to a $0\nu2\beta$ decay NME arises from the transformation of nucleon pairs that couple to total angular momentum $J = 0$~\cite{Caurier:08,fedor,iwata}. Such a decay corresponds to spherical superfluid parent and daughter nuclei. The $J \ne 0$ contributions to the matrix element arise from higher seniority~\cite{Caurier:08,fedor} components in the wavefunctions, due to broken Cooper pairs of nucleons. These lead to cancellations and effectively reduce the $0\nu2\beta$ decay amplitude. An additional reduction in the NME is expected if there is a seniority mismatch between the initial and the final wavefunctions~\cite{Caurier2008}. This would be the case if the parent and the daughter have  different intrinsic shapes~\cite{poves}, that are driven by multipole correlations. It is now established that these collective correlations (other than pairing) play an important role in $0\nu 2\beta$ NME calculations~\cite{Engel:2017} and could further quench calculated NMEs~\cite{poves}. As examples, one can look at two traditionally used many-body methods, the Interacting Shell Model (ISM) and the Quasiparticle Random Phase Approximation (QRPA). In the former, the treatment of correlations is exact, with comparatively smaller valence spaces. On the other hand, the QRPA calculations use larger model spaces, with relatively simpler configurations for the valence nucleons. Here, the pairing between like nucleons is treated via a transformation to the quasiparticle regime, within the Bardeen-Cooper-Schrieffer (BCS) approximation~\cite{avignone}. Such BCS pairing smears the Fermi surfaces in the parent and the daughter (as one would expect in a superfluid), while the RPA correlations admix higher seniority components in the wavefunctions~\cite{Caurier2008,fedor}. In this context, $(p,t)$~\cite{freeman,bloxham,thomas:12,sharp} and $(^{3}{\rm He},n)$~\cite{roberts} two-nucleon transfer experiments offer valuable insight into pairing correlations between like nucleons~\cite{broglia:73}. For reactions on even-even nuclei, strong population of the ground states (relative to excited $0^+$ states) would imply that both the target and residual nucleus ground states are nearly superfluid, and well described by BCS wavefunctions~\cite{freeman,broglia:73}.

For the case of $^{136}{\rm Xe}$ ${\to}$ $^{136}{\rm Ba}$, the $0\nu 2\beta$ decay NME differs more than a factor of four, depending on the many-body technique used~\cite{Rebeiro}. This is an important issue as $^{136}$Xe decay is a promising candidate to search for $0\nu2\beta$ decays~\cite{Rebeiro}. In fact, several planned large-scale time projection chamber (TPC)-based experiments aim to measure $^{136}$Xe $0\nu2\beta$ decay~\cite{nexo,lux,darwin,next}. From a nuclear structure perspective this is an interesting case, because of its location in a shape-transitional region of the Segr\`e chart. While the nearly spherical $^{136}$Xe has a closed shell at $N = 82$, the daughter $^{136}$Ba has $N = 80$. 

In light of the above, we recently benchmarked the $J = 0$ part of the $^{136}$Xe $0\nu2\beta$ decay NME, via a study of neutron pairing correlations using the $^{138}\mathrm{Ba}(p,t)$ reaction~\cite{Rebeiro}. Contrary to what one would expect for spherical superfluid systems~\cite{broglia:73}, we observed for the first time a strong population of higher lying $0^+$ states, relative to the ground state in $^{136}$Ba. About 53\% of the ground state $L = 0$ $(p,t)$ strength was distributed over excited $0^+$ states, with $\sim$~$35\%$ concentrated at the $0^+_2$, $0^+_3$ and $0^+_4$ levels. This was clear evidence of a breakdown of the BCS approximation for neutrons in $^{136}$Ba. The results also implied significantly different ground state wavefunctions for the spherical $^{138}$Ba and the (final-state) $^{136}$Ba nucleus. If this were the case, because one can expect nearly identical ground state wavefunctions for the $N = 82$ $^{138}$Ba and $^{136}$Xe nuclei, a sizable difference in deformation (seniority mismatch) between the parent and the daughter in $^{136}$Xe $\to$ $^{136}$Ba $\beta\beta$ decay cannot be ruled out. 

Further investigations of shape-transitional barium isotopes around $A \sim 130$ are relevant in context of the above. Previous data from $(p,t)$ reactions on barium isotopes show inconclusive evidence in this regard. For the $^{136}\mathrm{Ba}(p,t)$ case, C\v{a}ta-Danil \textit{et al.}~\cite{cata} observe around $34\%$ of the ground state strength distributed over excited $0^+$ states in $^{134}$Ba. However their measured $L = 0$ strength distribution was around a factor of two larger than what was reported in subsequent work by Pascu~\textit{et al.}~\cite{pascu1}. Similarly, for the $^{134}\mathrm{Ba}(p,t)$ reaction, Ref.~\cite{cata} report $\sim 27\%$ of the ground state strength distributed over excited $0^+$ states, while Ref.~\cite{pascu1} claim this value is much smaller, at around $10\%$.
Adequate $^{136}{\rm Ba}(p,t)$ angular distribution data were not presented in both references.
In comparison, measured $^{132,130}\mathrm{Ba}(p,t)$ cross sections do not show a similar fragmentation of the monopole strength~\cite{pascu2,suliman}.     
\begin{figure}[t]
\includegraphics[scale = 0.3]{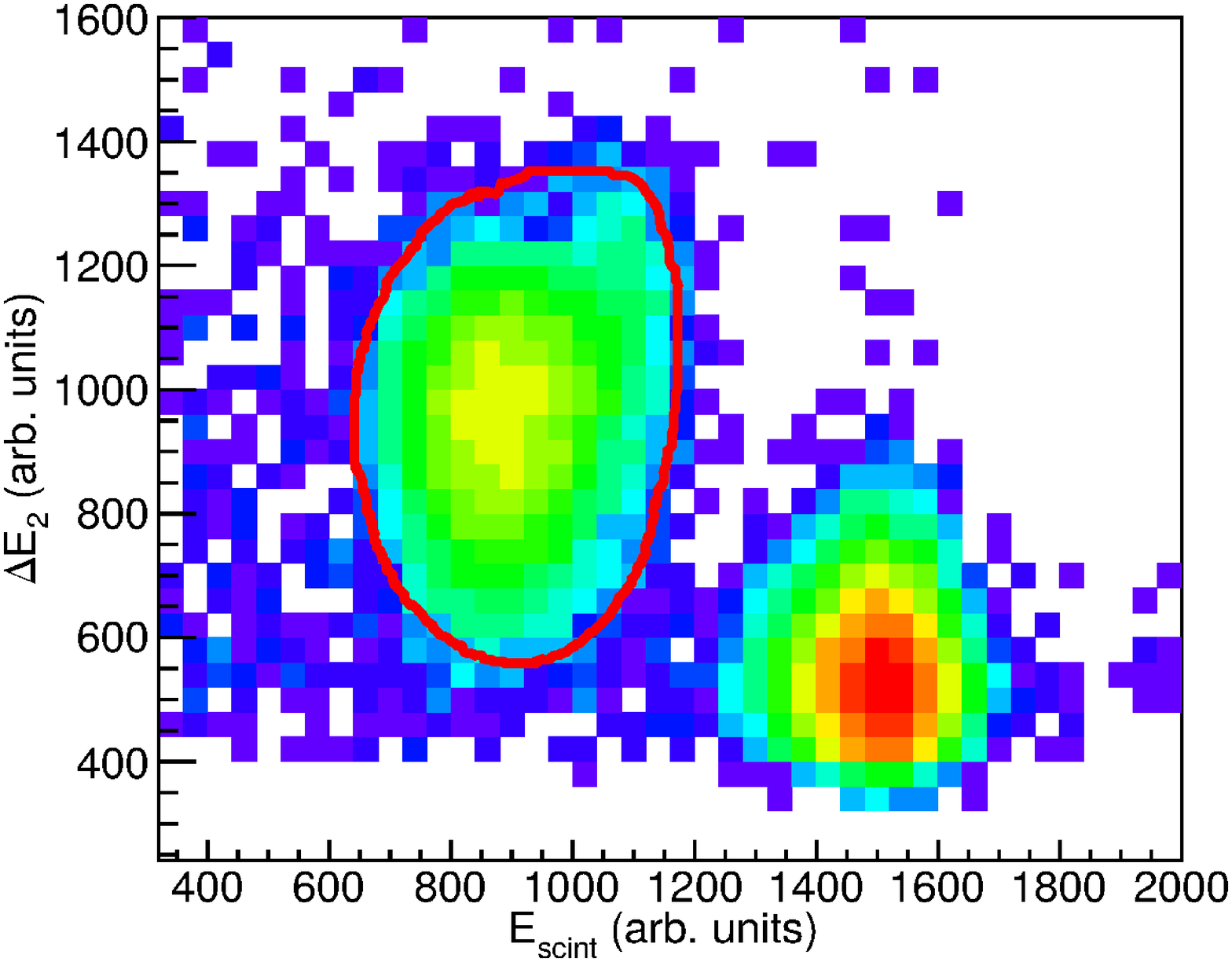}\\
\vspace{2em}
\includegraphics[scale = 0.31]{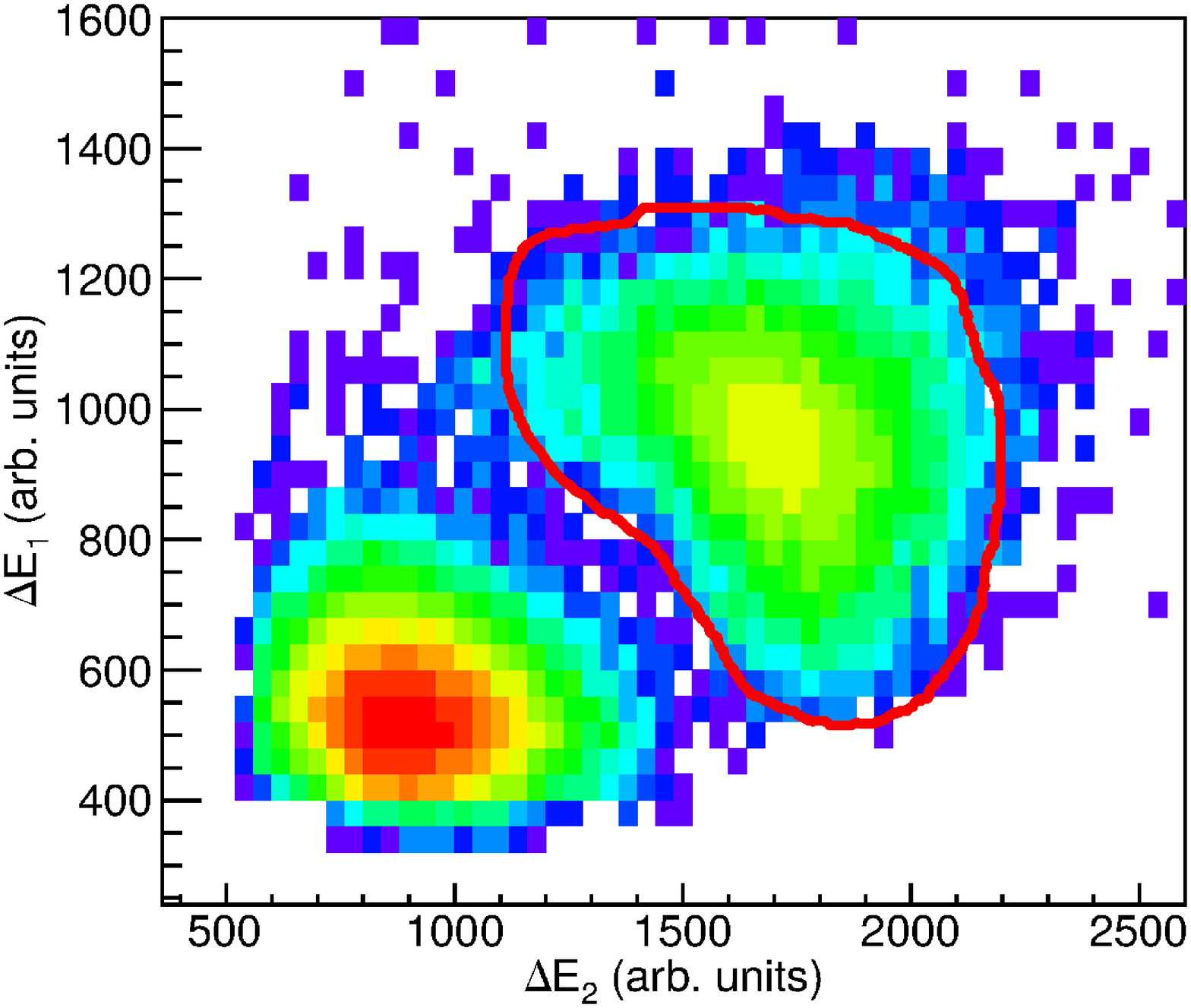}
\caption{\label{fig:PID}Particle identification spectra using energy losses registered in the two proportional counters and the total energy deposited in the plastic scintillator. The triton groups are highlighted. The other particles are mainly deuterons.}
\end{figure}

Considering the aforementioned discrepancies and given the importance of the measured results, in this Letter 
we report remeasurements of the relative population of $0^+$ states in $^{134}$Ba, with the $^{136}\mathrm{Ba}(p,t)$ reaction. The experiment was performed at the Maier-Leibnitz-Laboratorium~(MLL), operated jointly by the Ludwig-Maximilian Universit\"{a}t (LMU) and Technische Universit\"{a}t M\"{u}nchen (TUM), in Garching, Germany. 22~MeV protons from the MLL tandem accelerator were directed onto a 40~$\mu g/{\rm cm}^2$, 93$\%$ isotopically enriched $^{136}\rm{BaO}$ target, that was evaporated on a thin carbon foil. The light reaction products were momentum analyzed with the high-resolution Q3D magnetic spectrograph~\cite{Loffler:1973,Thomas}. The focal plane detector for the spectrograph comprised two gas proportional counters 
and a 7-mm-thick plastic scintillator~\cite{Thomas}. The energy losses of the charged ejectiles in the proportional counters and the residual energy deposited in the plastic scintillator were used to discriminate the tritons from other ejectiles (as shown in Fig.~\ref{fig:PID}).
A cathode strip foil in the second proportional counter provided high-resolution position information for the tritons.  
\begin{figure}[t]
\includegraphics[scale = 0.41]{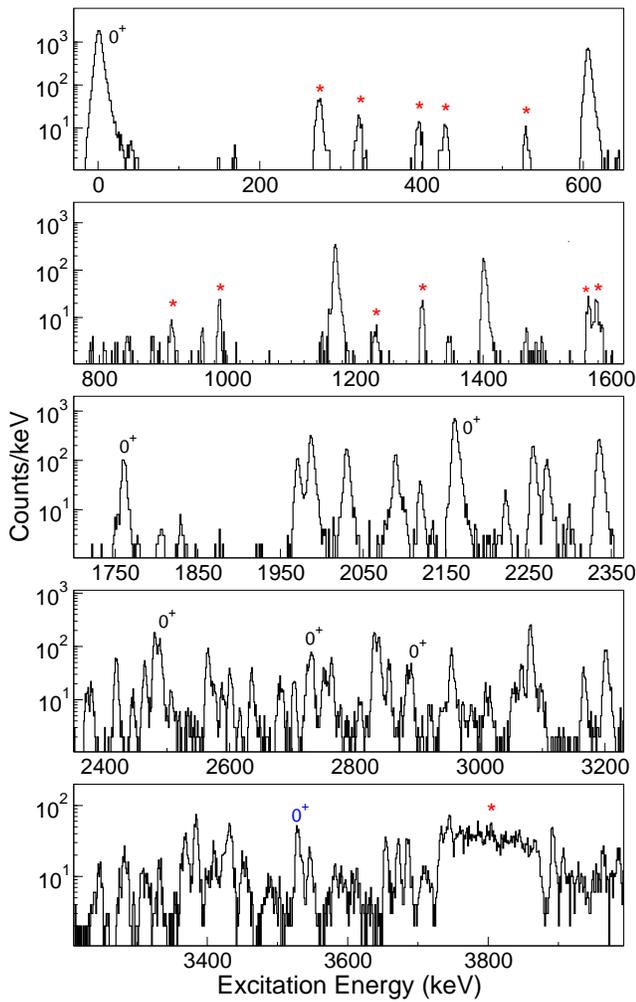}
\caption{{\label{triton_spectrum}} Triton spectra corresponding to states in $^{134}$Ba, at $\theta_{lab}$ = 25$^{\circ}$. Peaks from 
contaminants in the target are marked with asterisks.
The $0^+$ states observed in this experiment are labeled. The state at $\sim$3.5~MeV (labeled in blue) is reported for the first time in this work.}
\end{figure}
 \begin{figure}[t]
  \includegraphics[scale=0.32]{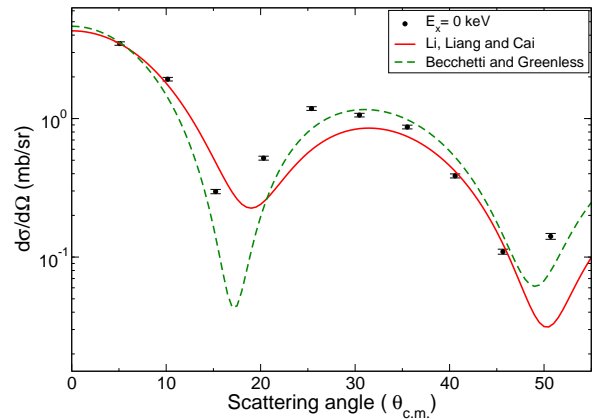}
\caption{{\label{fig:triton_omp}} Measured  $^{134}$Ba ground state angular distribution compared with normalized DWBA predictions obtained using different triton OMPs~\cite{Li,ripl1}.}
%
\end{figure}
\begin{figure}[t]
  \includegraphics[scale=0.34]{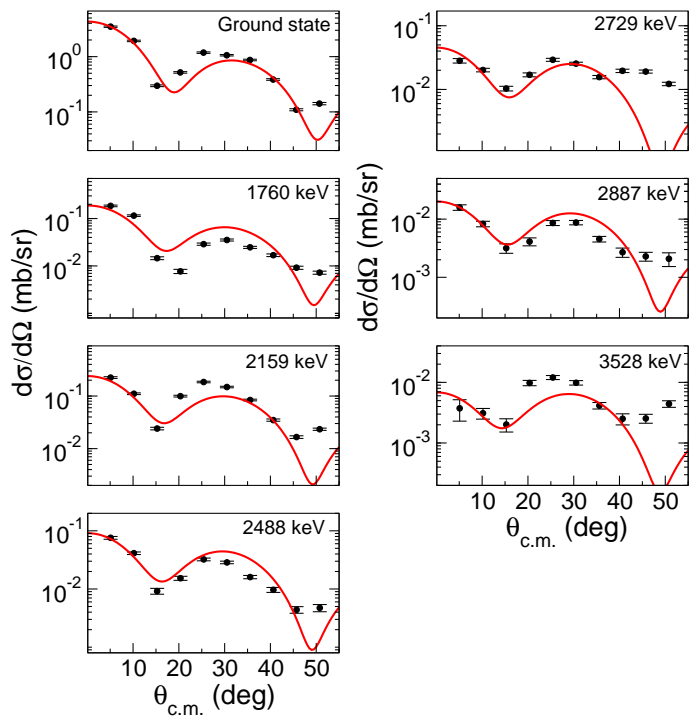}
\caption{{\label{cs_angular_distributions}} $^{136}\mathrm{Ba}(p,t)$ angular distributions for observed $0^+$ states in $^{134}\rm{Ba}$, compared with normalized DWBA predictions for $L = 0$ transfer.}
\end{figure}

We collected triton spectra at five magnetic field settings and at ten angles, ranging from 5$^{\circ}$ to 50$^{\circ}$ (in 5$^{\circ}$ increments). The Q3D field settings allowed us to study states in $^{134}$Ba up to $\sim$~4.0~MeV in excitation energy, with full-widths at half-maximum of $\lesssim$ 10~keV for the triton peaks. The triton energies were calibrated \textit{in-situ}, using well known states in $^{134}$Ba~\cite{nndc}. A sample spectrum is shown in Fig.~\ref{triton_spectrum}. 

During the course of the experiment, we also collected $^{136}$Ba($p,p$) data over an angular range of 10$^{\circ}$- 60$^{\circ}$. These data allowed us to accurately determine the effective $^{136}$Ba areal density in the target foil, for small angle scattering. In addition, both the $(p,p)$ and the $(p,t)$ data sets from this work were analyzed as described in Ref.~\cite{Rebeiro}, given the similarity between the two experiments. We performed distorted wave Born approximation (DWBA) calculations of angular distributions using the DWUCK4 code~\cite{DWUCK4} and compared them to our experimental results. The DWBA analysis used Woods-Saxon potentials and global optical model potential (OMP) parameters. Based on the agreement with our measured $^{136}\mathrm{Ba}(p,p)$ cross sections and a comparison with previous $^{138}\mathrm{Ba}(p,p)$ data obtained over a large angular range~\cite{Rebeiro}, we chose the global OMP parameters recommended by Varner {\it et~al.}~\cite{Varner} for the incoming proton channel in the DWUCK4 calculations. For the triton channel 
we chose the OMPs recommended by Li, Liang, and Cai~\cite{Li},  
as they yielded better agreement with our measured ground state angular distribution for $^{134}$Ba. This is shown in Fig.~\ref{fig:triton_omp}. The two-neutron transfer form factor was obtained using the OMPs from Ref.~\cite{neutronomp}, assuming a $(0h_{11/2})^2$ configuration. 
For each state, the depth of the potential was adjusted such that each transferred neutron had a binding energy $(S_{2n}+E_x)/2$, where $S_{2n}$ is the two neutron separation energy of $^{136}$Ba and $E_x$ is the excitation energy in the residual $^{134}$Ba nucleus. 

The above DWBA prescription was used to perform an angular distribution analysis of all the peaks shown in Fig.~\ref{triton_spectrum}, that corresponded to states in $^{134}$Ba. We identify seven $0^+$ states in $^{134}$Ba.
These include the ground state and a previously unreported level at 3528~keV.\footnote{An independent confirmation of this state would be welcome.} The measured angular distributions are shown in Fig.~\ref{cs_angular_distributions}. These data determined the monopole $(p,t)$ strengths to excited $0^+$ states, using the ratio~\cite{Rebeiro} 

\begin{equation}
 \epsilon_i = \left[
 \frac{ \left(\frac{d\sigma}{d\Omega}\right)^{\rm data}_{0^+{\rm ex}}}{\left(\frac{d\sigma}{d\Omega}\right)^{\rm DWBA}_{0^+{\rm ex}}}
 \right]_i 
 \left[
 \frac{ \left(\frac{d\sigma}{d\Omega}\right)^{\rm DWBA}_{{\rm G.S.}}}{\left(\frac{d\sigma}{d\Omega}\right)^{\rm data}_{{\rm G.S.}}}
 \right]~,
\end{equation}
so that the $Q$ value dependence on the cross sections was removed. 
\begin{table}[t]
\begin{flushleft}
\caption{
Measured $L = 0$ $^{136}\mathrm{Ba}(p,t)$ strength distribution over excited $0^+$ states in $^{134}$Ba, relative to the ground state.$^a$ For comparison we also list the results from previous work~\cite{cata,pascu1}. 
}
\label{tab:pt_strengths}
\begin{ruledtabular}
\begin{tabular}{l.....}
\multicolumn{2}{c}{Ref.~\cite{cata}}&\multicolumn{2}{c}{Ref.~\cite{pascu1}}&\multicolumn{2}{c}{This work}\\
\cline{1-2}\cline{3-4}\cline{5-6}
\multicolumn{1}{l}{$E_x$}&\multicolumn{1}{c}{$\epsilon_i$ }&\multicolumn{1}{c}{$E_x$}&\multicolumn{1}{c}{$\epsilon_i$}&\multicolumn{1}{c}{$E_x$}&\multicolumn{1}{c}{$\epsilon_i$}\\
\multicolumn{1}{l}{\rm (keV)}&\multicolumn{1}{c}{(\%)}&\multicolumn{1}{c}{\rm (keV)}&\multicolumn{1}{c}{(\%)}&\multicolumn{1}{c}{\rm (keV)}&\multicolumn{1}{c}{(\%)}\\
 \colrule
      1759 &3.73     &1760.8(3)      &2.6(1)  &1760.3(3)       & 10.3(4)  \\ 
      2161 &14.85    &2159.1(2)      &8.8(3)  &2159.6(3)        & 16.6(5)  \\
      2336 &\le 1.05 &...       & ...    &2334.2(3)        & \le 3.0       \\
      2378 &0.52     &2380.7(9)^b  & 0.10(1)&2379.0(4)        & \le 0.5        \\
      2485 &6.67     &2488.6(1)      & 2.2(1) &2488.4(3)       & 8.3(4)      \\ 
      2722 &1.99     &2727.5(2)      & 1.0(1) &2729.0(4)       & 5.5(3)      \\
      2874 &1.81     &2883.8(2)      & 0.8(1) &2887.0(5)       & 2.6(2)         \\
      ...  & ...     &2961.1(12)^b &0.10(1) &...        & ...          \\
      2996 &0.63     &3000.6(2)      &0.3(1)  &...        & ...           \\
      3181 &1.40     & ...      & ...    &...        & ...            \\
      ... &...       &3395.1(10)      &0.10(1) &...        & ...            \\
      3501 &1.47     &3505.4(5)      &0.5(1)  &...        & ...             \\
      ...  &...      &...       &...     &\multicolumn{1}{c}{3528(1)} & 1.6(2)  \\
      ...  & ...     &3602.3(11)^b  &0.10(1) &    ...    & ...              \\
      3618 &1.22     &3623.9(4)^b& 0.4(1) &...        & ...               \\
      ...  & ...     &3750.4(10)^b &0.2(1)  &...        &...                 \\

\colrule
\multicolumn{2}{c}{$\sum\epsilon_i$ = 34\%}& \multicolumn{2}{c}{$\sum\epsilon_i$ = 17.1(4)\%}&\multicolumn{2}{c}{$\sum\epsilon_i$ = 44.9(9)\%}
\end{tabular}
$^a$ The $0^+$ assignments for states above 2488~keV were only made from $^{136}\mathrm{Ba}(p,t)$ measurements.\\
$^b$ These states were tentatively assigned spin-parity $J^\pi = (0^+)$.
\\ 
\end{ruledtabular}
 \end{flushleft}
 \end{table}

The results are shown in Table~\ref{tab:pt_strengths}. We find that our extracted value for the \textit{integrated strength} is more consistent with the result of C\v{a}ta-Danil {\it et al.}~\cite{cata} and disagrees significantly with the later work of Pascu~{\it et al.}~\cite{pascu1}. It is also apparent that there are some large discrepancies between our work and Ref.~\cite{pascu1} for individual states, and that we
observe fewer excited $0^+$ states than either of these measurements.\footnote{The results of Refs.~\cite{cata,pascu1} are also discordant with one another, for both $^{136,134}\mathrm{Ba}(p,t)$ data.}  We briefly discuss a few aspects of this comparison below. 

The Nuclear Data Sheets (NDS) for $A = 134$~\cite{nds} list a doublet of states at 2334.76(6) and 2336.82(3)~keV. These levels are assigned spin-parity $(1,2^+)$ and $0^+$ respectively. 
The Q3D spectrograph (also used in Refs.~\cite{cata,pascu1}) is limited by energy resolution to differentiate between these two states. We observe a single triton peak corresponding to $E_x = 2334.2(3)$~keV, which has an angular distribution that is more consistent with an $L = 2$ transition.
\footnote{A detailed analysis of the full data set will be presented in a future publication.}
This indicates that the population of the 2337~keV level was below our experimental sensitivity. We encounter a similar situation near 2.4~MeV, with two known closely-spaced states at 
2379.112(18)~keV~\cite{nds} and 2377.1(4)~keV~\cite{Fazekas,nds}. The former is an established $0^+$  level, while the latter was assigned a tentative spin value of $J = (6)$~\cite{nds}.
Our spectrum shows a triton peak corresponding to 2379.0(4)~keV, whose measured angular distribution agrees well with $L = 2$ transfer. 

As a result of the above, we measured cross sections at 5$^\circ$ to place upper limits on the possible $L = 0$ strengths for the 2337 and 2379~keV levels.  
Several of the other weakly populated $0^+$ states at higher excitation energy were not observed in this work. A probable explanation for this can be invoked from the fact that the  previous measurements were carried out at a higher beam energy of 25~MeV.
Due to the large negative $Q$ value for this reaction ($-7.6$~MeV), the 
$L = 0$ cross section at the most forward angles decreases more rapidly with increase in excitation energy for 22~MeV protons as compared to 25~MeV. Our DWBA calculations show that at 22~MeV, the forward angle differential cross sections are around a factor of 2 smaller at 1.8~MeV excitation energy, compared to those at 25~MeV. This factor increases to approximately 5 and 10 at excitation energies of 2.5 and 3.5 MeV, respectively. In contrast, large changes in forward angle differential cross sections are not predicted for $L=2$ transitions. 

It is difficult to comment further since both Refs.~\cite{cata,pascu1} reported meager $^{136}\mathrm{Ba}(p,t)$ angular distribution data. C\v{a}ta-Danil {\it et al.}~\cite{cata} acquired data at only two angles and chose to identify the $L = 0$ transitions using a single number, the ratio of the cross sections at $\theta_{\rm lab} = 6^\circ$ and $15^\circ$. In comparison, Pascu \textit{et al.} acquired data at only three angles~\cite{pascu1} and do not explicitly show angular distributions for excited $0^+$ states. They identified the $0^+$ states using the same procedure as in Ref.~\cite{cata}. Adequate descriptions regarding the choice of OMP parameters for their analyses and the procedure to determine target thicknesses 
were not provided in these references. However, since both these works provided more complete $^{134}{\rm Ba}(p,t)$ angular distribution plots, we are able to address the discrepancy for this reaction.  In particular, our estimated values of $\epsilon_i$ (via inspection and from DWBA calculations) show better agreement with the results of Ref.~\cite{cata} and significantly disagree with the results in Ref.~\cite{pascu1}. This gives credence to the former results over the latter.     

It is evident from our results in Table~\ref{tab:pt_strengths} that a significant portion of the $(p,t)$ strength is distributed over excited $0^+$ levels in $^{134}$Ba. This is similar to our previously reported $^{138}\mathrm{Ba}(p,t)$ results~\cite{Rebeiro} and clearly indicates a breakdown of the BCS approximation for neutrons in $^{134}$Ba. The persistence of such behavior
as one moves away from the $N = 82$ shell closure, with integrated $^{138,136,134}{\rm Ba}(p,t)$ strengths being $\sim$ 53\%, 45\% and 27\% highlights the shape-transitional nature of these isotopes. This implies significantly different deformations in the ground states of $^{138}$Ba, $^{136}$Ba and $^{134}$Ba and a mismatch between the wavefunctions for the initial and final states in $^{136}$Xe $\beta\beta$ decay. Such a curtailed overlap
offers a possible explanation for the comparatively long $2\nu2\beta$ decay half-life~\cite{ruben} measured for $^{136}$Xe. It would also reduce its calculated $0\nu2\beta$ decay NME compared to a scenario in which both the parent and daughter are nearly spherical (or similarly deformed). Both experimental investigations of quadrupole correlations in $^{136}$Ba as well as theoretical studies of deformation effects on the NME (along the lines of Refs.~\cite{rath,fang}) will be useful to further shed light in this regard. 

This work was partially funded by the National Research Foundation (NRF), South Africa under Grant No. 85100. J.C.N.O thanks the NRF funded MaNuS/MatSci
program at UWC for financial support during the course of his M.Sc. 

\bibliography{136Ba_pt}

\end{document}